\documentclass[floatfix,aps,prl,twocolumn,superscriptaddress,showpacs]{revtex4-1}
\usepackage[normalem]{ulem}
\usepackage{float}
\usepackage[usenames]{color}
\usepackage{graphicx}
\usepackage{ulem}
\usepackage{epstopdf}
\usepackage{amsfonts}
\usepackage{amsmath}

\begin{document}

\title{Broken Detailed Balance of Filament Dynamics in Active Networks}

\author{J. Gladrow}
\email{Present address: Cavendish Laboratory, Cambridge, UK.}
\affiliation{Third Institute of Physics, Georg August University, 37077 G\"{o}ttingen, Germany }
\author{N. Fakhri}
\affiliation{Physics of Living Systems Group, Department of Physics, Massachusetts Institute of Technology, Cambridge, MA 02139, USA}
\affiliation{Kavli Institute for Theoretical Physics, University of California, Santa Barbara, California 93106, USA}
\author{F.C. MacKintosh}
\affiliation{Department of Physics and Astronomy, Vrije Universiteit, 1081 HV Amsterdam, Netherlands}
\affiliation{Kavli Institute for Theoretical Physics, University of California, Santa Barbara, California 93106, USA}
\author{C.F. Schmidt}
\email{christoph.schmidt@phys.uni-goettingen.de}
\affiliation{Third Institute of Physics, Georg August University, 37077 G\"{o}ttingen, Germany }
\affiliation{Kavli Institute for Theoretical Physics, University of California, Santa Barbara, California 93106, USA}
\author{C.P. Broedersz}
\email{C.broedersz@lmu.de}
\affiliation{Arnold-Sommerfeld-Center for Theoretical Physics and Center for
  NanoScience, Ludwig-Maximilians-Universit\"at M\"unchen,
   D-80333 M\"unchen, Germany.}
\affiliation{Kavli Institute for Theoretical Physics, University of California, Santa Barbara, California 93106, USA}

\date{\today}

\begin{abstract}
Myosin motor proteins drive vigorous steady-state fluctuations in the actin cytoskeleton of cells. Endogenous embedded semiflexible filaments such as microtubules, or added filaments such as single-walled carbon nanotubes are used as novel tools to non-invasively track equilibrium and nonequilibrium fluctuations in such biopolymer networks. Here we analytically calculate shape fluctuations of semiflexible probe filaments in a viscoelastic environment, driven out of equilibrium by motor activity. Transverse bending fluctuations of the probe filaments can be decomposed into dynamic normal modes. We find that these modes no longer evolve independently under nonequilibrium driving. This effective mode coupling results in non-zero circulatory currents in a conformational phase space, reflecting a violation of detailed balance. We present predictions for the characteristic frequencies associated with these currents and investigate how the temporal signatures of motor activity determine mode correlations, which we find to be consistent with recent experiments on microtubules embedded in cytoskeletal networks.

\end{abstract}
\maketitle
\noindent Living cells are a prime example of active matter, typically driven out of thermodynamic equilibrium by molecular force generators~\cite{CliffJCB08, fletcherMullins,Schmidt10,Marchetti, prost2015active}. In cytoskeletal networks, myosin motors are the dominant actors driving nonequilibrium dynamics~\cite{Howard, lau2003microrheology,Mizuno09,FakhriHighResolutionMapping, Guo2014Cell,Weber2015,Battle2016,Rupprecht}, which can be captured in simplified reconstituted systems~\cite{Mizuno2007, BrangwynneNonequilibriumMicrotubule, polarPatterns,frozenStates,Loi2011, murrel2012, molecularMotorsRobustlyDrive}. Myosin motors generate mechanical force by coupling the hydrolysis of adenosine triphosphate (ATP) to conformational changes in a mechano-chemical cycle~\cite{Howard}. Recently, it was demonstrated that motor activity can break detailed balance, giving rise to circulating probability currents in a phase space of collective degrees of freedom in cellular systems~\cite{Battle2016}. However, for cytoskeletal networks how broken detailed balance propagates from molecular-scale dynamics to large-scale collective dynamics remains unclear. An elegant way of non-invasively investigating the dynamics of active biological networks across a range of length scales is to track embedded probe filaments such as  microtubules~\cite{BrangwynneNonequilibriumMicrotubule} or single-walled carbon nanotubes~\cite{FakhriHighResolutionMapping}. A careful analysis of the shape fluctuations of such a probe filament can reveal the active nature of its environment including potential violations of detailed balance, but a theory for the stochastic dynamics of these systems is still lacking.

\begin{figure}
  \begin{center}
    \includegraphics[width=0.95 \columnwidth]{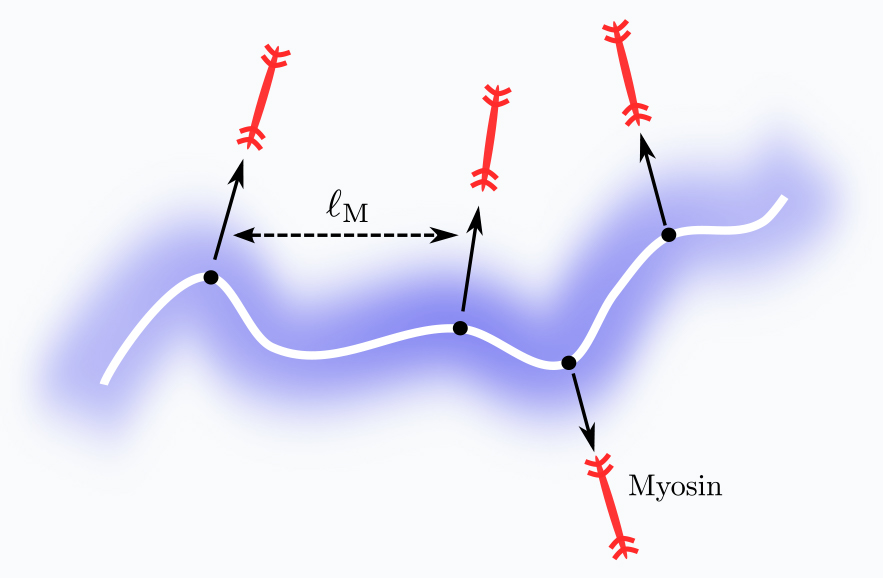}
    \vspace{-0.1in}
    \caption{(color online) Schematic of the model: A semiflexible probe filament (blue) is embedded in a crosslinked actin-myosin network (light grey). Actin filaments in the network are contracted by engaged myosin motors (red). This active network interacts with the probe filament through entanglement points (black dots) spaced at a length scale $\ell_\text{M}$, inducing nonequilibrium bending fluctuations in the filament.}\label{fig:networks}
  \end{center}
\end{figure}

Here, we develop an analytical theory describing the nonequilibrium dynamics of a large probe filament embedded in an active viscoelastic gel as a model for a cytoskeletal network with spatially distributed motors (Fig.~\ref{fig:networks}). We derive a Fokker-Planck  description of the statistical properties in a phase space spanned by the bending modes of the probe filament. Under steady-state conditions, we find broken detailed balance in the form of nonzero probability currents, which exhibit rotations in phase space  with characteristic frequencies. Conceptually, the breakdown of detailed balance in mode space is found to arise from a nonuniform  distribution of the active forces acting on the filaments, where sites that experience higher motor activity dissipate mechanical energy along the probe filament. These currents are a telltale signature of nonequilibrium dynamics and can in principle be inferred directly from experiments~\cite{Battle2016}, providing a noninvasive method to identify and quantify nonequilibrium dynamics. We consider the appropriate dynamic normal modes of the probe filaments. These modes evolve independently in thermodynamic equilibrium. By contrast, we find that motor activity  induces correlations between the dynamic normal modes. Furthermore activity selectively enhances bending fluctuations on a characteristic length scale that stems from the balance between filament bending elasticity and network shear elasticity, as observed in experiments~\cite{BrangwynneNonequilibriumMicrotubule}.

Building on previous work~\cite{MacKintosh2008, lau2003microrheology, Liverpool2003}, we propose a model for the fluctuations of a semiflexible probe filament of length $L$ embedded in a homogeneous viscoelastic network at temperature $T$. Randomly distributed motors are added to this gel and exert forces on the filaments (e.g. actin filaments) forming the network. Stresses then propagate through the gel and act on the probe filament at entanglement points between this filament and the network (Fig.~\ref{fig:networks}). We assume that both thermal forces and motor forces propagated through the network induce small transverse fluctuations of the probe filament. To capture these fluctuations, we parameterize the shape of the weakly undulating filament at time $t$ by the transverse deflection $r_{\perp}(s,t)$ along its arc length $s$. Such transverse deflections result in a restoring force per unit length $f_{\rm bend}=- \kappa \partial^4 r_\perp/\partial s^4\,(s,t)$, assuming the filament can be described as an inextensible worm-like chain~\cite{aragon1985dynamics, kratkyPorod, chaseReview} with a bending rigidity $\kappa$. When the  filament bends, the resulting network deformations also exert forces on the filament, 
$f_{\rm network}=\int_{-\infty}^t \mathrm{d}t' \alpha(t~-~t')r_\perp(s,t')$, where the memory kernel $\alpha(t)$ represents the response function to a transverse force, capturing the viscoelastic properties of the embedding medium. 

The probe filament is entangled with the network at  points separated by a length, $\ell_\text{M}$, of the order of the mesh-size. Thus, we will assume that active forces with an average amplitude $f_i$ act independently on the probe filament at discrete sites $s_i$, with a typical separation $\ell_\text{M}$ along the filament (Fig.~\ref{fig:networks}). We do this to model the dominant effect of nearby motors in the network, and we  do not include far-field contributions from distant motors~\cite{Toyota2011, Gov2015}. Under these assumptions, the total motor-induced forces on the probe filament can be approximated as 
\begin{align}
  \label{eq:motorForce}
  f_\text{M}(s,t) \approx \sum\limits_{i}\, f_i \mathcal{T}_i(t)\delta\left(s-s_i \right).
\end{align}
We model the dynamics of active forces, $f_i \mathcal{T}_i(t)$, at the point where they act on the probe filament as independent random step-like processes with a constant amplitude~$\lvert f_i \rvert$. For simplicity, we ignore fluctuations in the active force amplitude. In other words, the temporal behavior of a single motor is assumed to be a telegraph process: $\mathcal{T}(t)$ switches randomly from zero (motor not engaged) to one (motor engaged) at a rate $\tau_\text{on}^{-1}$, and back to zero at a rate $\tau_\text{off}^{-1}$. The autocorrelation of the motor forces is then~\cite{gardiner}
\begin{align}
  \label{eq:telegraphCorrelation}
  \langle \mathcal{T}_n(t)\mathcal{T}_n(t') \rangle = C_1+C_2 e^{-\frac{|t-t'|}{\tau_{\rm M}}} ,
\end{align}
where $\tau_\text{M}^{-1}=\tau_\text{on}^{-1}+\tau_\text{off}^{-1}$, and $C_1~=~\tau_{\rm off}^2/\left(\tau_{\rm on}+\tau_{\rm off}\right)^2$ and $C_2~=~\tau_{\rm on}\tau_{\rm off}/\left(\tau_{\rm on}+\tau_{\rm off}\right)^2$ are dimensionless constants. Although this is a simple model for the dynamics of motor-generated forces, the corresponding power spectrum  $S(\omega)$ is Lorentzian. Such a Lorentzian spectrum captures the essential features observed in experiments \cite{lau2003microrheology, Guo2014Cell, Mizuno09, FakhriHighResolutionMapping} of becoming white-noise-like, $S(\omega)\sim \text{const}$, for low frequencies, while following a power-law $S(\omega)\propto \omega^{-2}$, for high frequencies. 

It is convenient to expand the deflections of the probe filament $r_\perp(s,t)=L \sum_q a_q(t)y_q(s)$ into a sum of orthogonal eigenmodes $y_q(s)$ with eigenvector $q$ of the beam equation $\gamma \partial r_\perp /\partial t = -\kappa \partial^4 r_\perp/\partial s^4$ with free boundary conditions at the ends \cite{aragon1985dynamics}. In the  $y_q(s)$-mode space the equation of motion of the probe filament becomes
\begin{align}
   \int \limits^t \mathrm{d}t'\, \alpha(t-t') a_q(t') &= -\kappa q^4 a_q(t) +f_{\text{M},q}(t)+\xi_q(t). \label{eq:modeEquationOfMotion}
\end{align}
This equation describes the force balance $f_{\rm network} =f_{\rm bend}+f_M+\xi$, with thermal forces $\xi$. Indexed quantities in Eq.~\eqref{eq:modeEquationOfMotion} denote projected variables, such as the  projected motor-induced force $f_{\text{M},q}(t)=\sum_{i} f_i/L \,y_q(s_i)\mathcal{T}_i(t)$. 
%\jannes{In all equations, we label modes according to their wave vector $q$ rather than their mode number $k$, as the latter depends on the boundary conditions, e.g. $q_{\text{fixed-ends}}(k) =\pi k/L$.}

%Note, thermal forces of differing modes do not correlate $\langle \xi^\text{Th}_k \xi^\text{Th}_j \rangle \propto \delta_{k,j}$, due to orthogonality of the eigenmodes~$y_k(s)$.

 In reconstituted actin networks and live cells, myosin activity has been found to lead to enhanced fluctuations of the bending modes of embedded microtubules~\cite{BrangwynneNonequilibriumMicrotubule, FakhriHighResolutionMapping}. Motivated by these experiments, we derived analytical predictions of mode fluctuations using realistic values for the time scales of myosin activity. The Fourier transform of the memory kernel $\hat{\alpha}(\omega)$ (Eq.~\eqref{eq:modeEquationOfMotion}) is related to the  shear modulus $G(\omega)$ of the medium through $ \hat{\alpha}(\omega)= k_0\,G(\omega)$, with a geometrical constant $k_0$~\cite{GittesMicroViscoelasticity, lamb1993hydrodynamics}. Here we use $k_0\approx 4\pi$, with logarithmic corrections depending  on the filaments dimensions~\cite{Howard}. For low enough frequencies, the viscoelastic properties of crosslinked networks of semiflexible polymers can be approximated as a simple elastic solid in a solvent with $G(\omega)\approx G_0-i\eta \omega$, where $\eta$ is the solvent viscosity. In general, the rheology of crosslinked actin networks can be characterized by two frequency regimes~\cite{ PhysRevLett.96.138307}: a low-frequency plateau regime with modulus $G_0$, and a high-frequency regime where the complex shear modulus scales as $G(\omega)\sim (-i \omega)^{3/4}$ \cite{MorseViscoelasticity, PhysRevE.58.R1241}. Here, we do not consider the latter, which typically sets in at frequencies of order $100$ Hz \cite{PhysRevLett.96.138307}, beyond frequencies where motor-generated fluctuations typically dominate~\cite{lau2003microrheology,Mizuno09,Guo2014Cell}.

Using this low-frequency model for the coupling between the probe filament and the active gel, we arrive at a mode correlation function, including thermal and motor-induced contributions, 
  $\langle a_q(t)a_w(t')\rangle~=~\langle a_q(t)a_w(t')\rangle_{\text{Th}}~+~\langle a_q(t)a_w(t')\rangle_{\text{M}}$
given by,
\begin{align}
  \langle a_q(t)a_w(t')\rangle_{\text{Th}} &=  \frac{k_B T \tau_q}{L^2\gamma}\delta_{q,w}e^{-\frac{\left | t-t'\right | }{\tau_q}}  \label{eq:thermalModeCorrelation},\\
  \langle a_q(t)a_w(t')\rangle_{\text{M}} &=    \frac{C_2}{L^2 \gamma^2}\mathbf{F}_{q,w}\mathcal{C}_{q,w}\left( t-t'\right) \label{eq:motorModeCorrelation},
\end{align}
 with coupling coefficients defined as $ \mathbf{F}_{q,w}= f^2\sum_{i}\, y_q(s_i)y_w(s_i)$ (see SI), and a mode relaxation time $\tau_q= \left(\kappa q^4/\gamma + G_0/\eta\right)^{-1}$, where $\gamma\approx4\pi \eta$. 
 
 \begin{figure}
  \centering
  \includegraphics[width=0.95 \columnwidth]{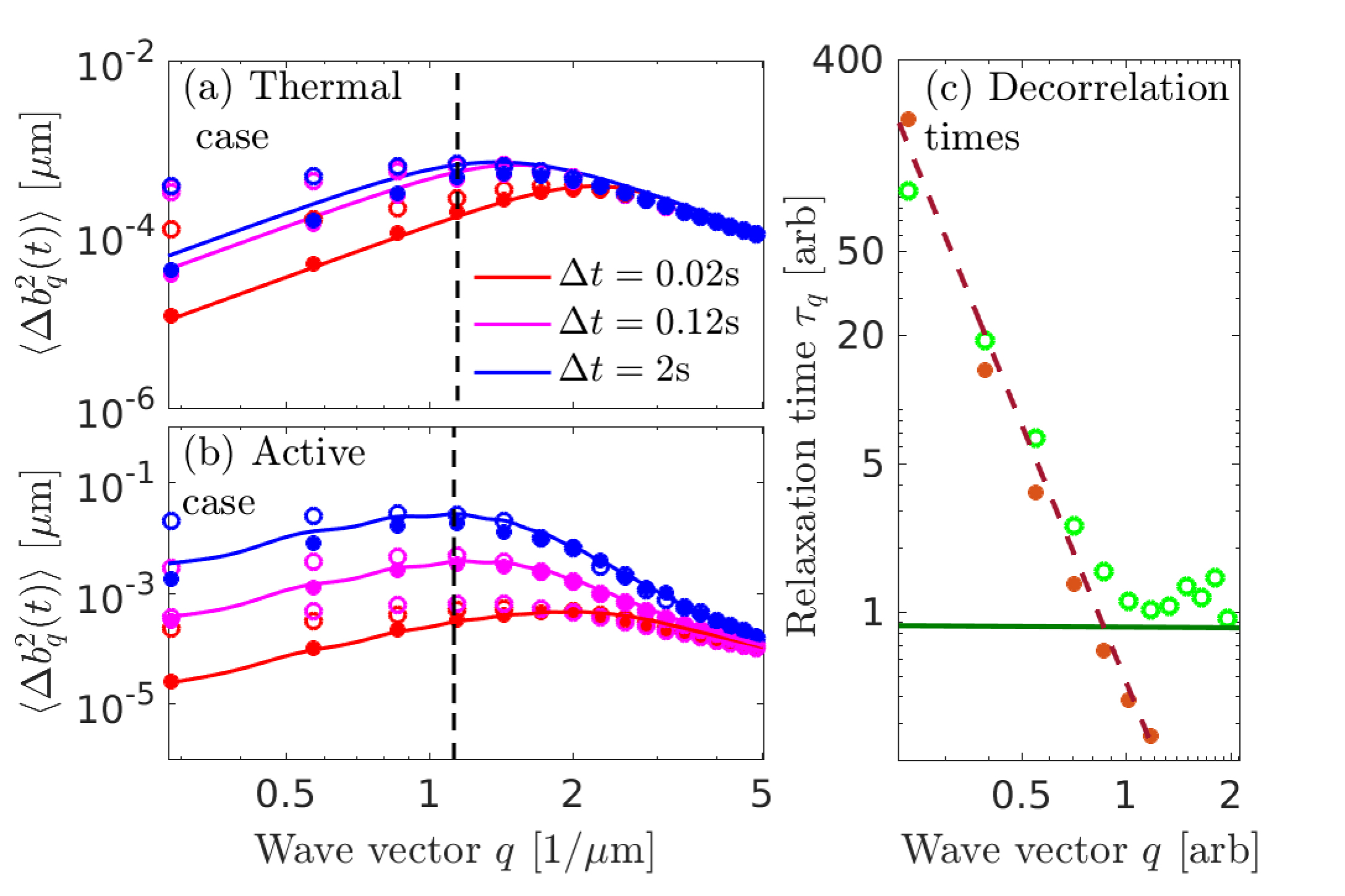}
  \caption{(color online) (a) Thermal $\theta$-mode fluctuations $\langle \Delta b_q^2(\Delta t)\rangle$ vs. wave vector $q$ for different lag times $\Delta$. Open symbols are values obtained from simulations with free-rod boundary conditions, while closed symbols are results for fixed ends. Continuous lines represent theoretical predictions. The vertical dashed line indicates $q^\ast$ as defined in the text. The viscosity and bending rigidity are given by  $ \eta=1 \,\text{Pa}\text{s}$  and  $\kappa=4\times 10^{-24} \,\text{N}\text{m}^2$(as in \cite{BrangwynneNonequilibriumMicrotubule}), while $L$, $T$,  $G_0$ and $\tau_\text{M}$ were set to $10 \, \mathrm{\mu m}$, $300$ K,  $10$ Pa and $0.5$ s respectively. (b) Active case at the same lag parameters and times as in (a) with $f=5$ pN and $\ell_\text{M}=0.45 \,\mu\mathrm{m}$. (c) Decorrelation times inferred from simulations of actively driven filaments (open symbols) converge to the motor time scale $\tau_\text{M}$ (green line). In the passive case, mode decorrelation times correspond to mode relaxation times which decrease (without bound) as $q^{-4}$ (dashed line).}
  \label{fig:fluctuationOverModeNumber}
\end{figure}

The function $\mathcal{C}_{q,w}(\Delta t)$ encodes the active contribution to the temporal evolution of the mode correlation function, which is characterized by a competition of mode and motor time scales. Regarding the autocorrelation ($\mathcal{C}_{q,q}(\Delta t)\equiv  \mathcal{C}_q(\Delta t)$), we obtain
\begin{align}
  \label{eq:4}
    \mathcal{C}_{q}(\Delta t) = \frac{1}{\tau_q^{-2}-\tau_\text{M}^{-2}}\left(e^{-\frac{\left | \Delta t\right|}{\tau_\text{M}}}-\frac{\tau_q}{\tau_\text{M}}e^{-\frac{\left|\Delta t\right|}{\tau_q}}\right).
\end{align}
The dynamics of filaments are conceptually similar to the active dynamics of other extended objects such as membranes~\cite{Garcia2015,Turlier2016}. Importantly,  $\mathcal{C}_{q}(\Delta t)$ in Eq.~\eqref{eq:4} is finite for all $q$, even when $\tau_q = \tau_\text{M}$. Naturally, the motor time scale $\tau_\text{M}$ imposes a lower bound on the relaxation times $\tau_q$, since modes can never decorrelate faster than the force that is driving them, consistent with simulations of active polymers~\cite{Ghosh20141065}. Indeed,  $\lim\limits_{_q \to \infty} \mathcal{C}_q(\Delta t)/\mathcal{C} _q(0) = e^{-\frac{|\Delta t|}{\tau_\text{M}}}$, which implies  that relaxation times inferred from mode fluctuations will converge to the motor time scale for high enough mode number. We verify this implication with Brownian dynamics simulations of filaments in active viscoelastic media, as shown in Fig.~\ref{fig:fluctuationOverModeNumber}(c).
To compare our theoretical predictions (Eqs.~(\ref{eq:thermalModeCorrelation}) and~(\ref{eq:motorModeCorrelation})) with experimentally recorded fluctuations, it is convenient to consider the backbone angle $\theta(s,t)\approx\partial r_\perp(s,t)/\partial s$ instead of the lateral deflection $r_\perp(s,t)$ \cite{GittesFlexuralRigidity}; $\theta$-mode amplitudes  $b_q$ are related to the $r_\perp$-modes through $b_q(t) = Lq\, a_q(t)$.
The predicted $\theta$-mode fluctuations,  $\langle \Delta b_q^2(\Delta t) \rangle =2\langle b_q^2(t)\rangle -2\langle b_q(t+\Delta t)b_q(t)\rangle $, exhibit enhanced fluctuations near the characteristic wave vector $q^\ast\sim (G_0/\kappa)^{1/4}$, as shown in Figs.~\ref{fig:fluctuationOverModeNumber}a,b.
This scaling law was derived for an infinite rod in a purely elastic medium \cite{landauLifschitz} and has been both discussed and experimentally observed by Brangwynne et al. \cite{BrangwynneNonequilibriumMicrotubule}. Here we recover this characteristic length scale as the scale where active fluctuations calculated in our analytic model reach a maximum.

\begin{figure}
   \centering
   \includegraphics[width= 0.95 \columnwidth]{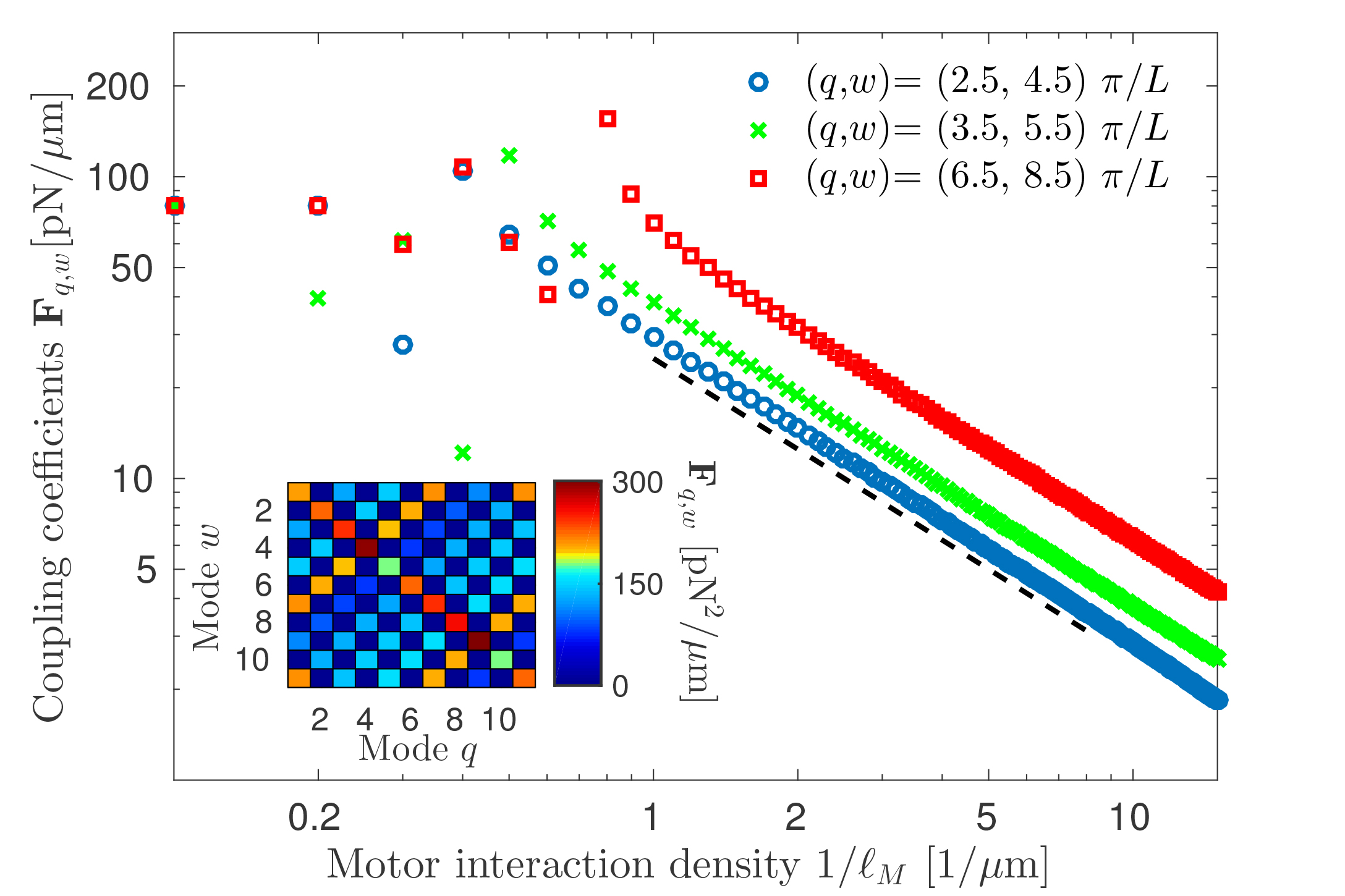}
   \caption{(color online) Coupling coefficients $ \mathbf{F}_{q,w}$ (defined in the text)  for three different mode pairs as a function of the motor interaction density $1/\ell_\text{M}$ with $f=10$ pN, $L=10\,\mu$m. Evenly spaced interaction sites $s_n$ were placed along the probe filament. Due to the orthogonality of the bending modes $y_q(s)$, the coupling strength decreases with density $1/\ell_\mathrm{M}$ (black line indicates power-law with exponent -1). The inset shows the coupling matrix for the first 11 modes for fixed interaction spacing ($\ell_\text{M}=1.66\, \mu $m).}
   \label{fig:correlationCoefficients}
 \end{figure}
 
We now consider the low-frequency dynamics of the probe filament, i.e. fluctuations at frequencies small compared to the rate of motor-driven events $\omega<2\pi/\tau_\text{M}$. At such low frequencies, motor correlation times become negligible, allowing us to approximate the telegraph process $\mathcal{T}_n( t)$ (see Eq.~\eqref{eq:telegraphCorrelation}) as white noise. 
Thus, the low-frequency limit enables us to describe the noise terms in Eq.~\eqref{eq:modeEquationOfMotion} by simple statistical processes, such that both thermal $\xi_q(t)$ and motor-induced processes $f_{\text{M},q}(t)$ can be combined in a single Gaussian random variable $\psi_q(t)$, with a correlator $\langle \psi_q(t)\psi_w(t')\rangle =2 \gamma^2  \mathbf{D}_{q,w}  \delta \left( t-t'\right)$, where the diffusion matrix is given by
\begin{align}
 \mathbf{D}_{q,w}&=  \frac{1}{2L^{2} \gamma^{2} }\left(2k_BT\gamma\delta_{q,w} + C_2\tau_\text{M} \mathbf{F}_{q,w} \right)  \label{eq:DiffusionMatrix}.
 \end{align}
An example of the matrix of coupling coefficients $ \mathbf{F}_{q,w}$ (see also Eq.~\eqref{eq:motorModeCorrelation}) is shown in the inset of Fig.~\ref{fig:correlationCoefficients} with a regular distribution of motor interaction points $s_i$. In this case, modes of different parity do not couple due to symmetry, while, for instance, modes located on the $7-7$ off-diagonal couple strongly in this example. Note, an uneven local distribution of active forces along the probe filament in a disordered network will in general lead to coupling between even and odd modes (See SI). We also find that the coupling strength decreases with increasing density of motor interaction points $\ell_\text{M}^{-1}$ (Fig.~\ref{fig:correlationCoefficients}). The coupling coefficients must indeed vanish for high motor density, since in this limit the  coupling coefficients represent an inner product of two orthogonal modes. 

%\chase{Furthermore, when performing an ensemble average over all possible locations of the probe filament in the network, the distribution of active forces along the filament will converge to a uniform distribution, resembling the high motor density limit, again with no effective coupling between modes.}

The nondiagonal structure of the diffusion matrix (Eq.~(\ref{eq:DiffusionMatrix})) has important implications for the nonequilibrium dynamics of the system. We explore these implications by developing a Fokker-Planck description of the dynamics of the probability density, $\rho(\vec{ a},t)$, where $\vec{ a}$ is the bending mode amplitude vector. This probability density satisfies the continuity equation 
$\frac{\partial \rho}{\partial t}= -\vec{\nabla}\cdot \vec{ j}$, where the current density $\vec{j}$ is given by

\begin{align}
 \vec{j}(\vec{a},t)=\mathbf{K}\vec{a}\rho\left(\vec{a},t\right)-\mathbf{D}\vec{\nabla}\rho\left(\vec{a},t\right).
\label{eq:current}
\end{align}
Here, $\mathbf{K}$ and $\mathbf{D}$ denote the deterministic matrix $\mathbf{K}_{j,k}=-\left( \kappa q_j^4/\gamma +G_0/\eta\right) \delta_{j,k}$, and 
the diffusion matrix (Eq.~\eqref{eq:DiffusionMatrix}). In general, broken detailed balance, i.e. finite $\vec{j}(\vec{a},t)$, is ensured when $ \left (\mathbf{K}\mathbf{D}\right)_{q,w}\neq \left(\mathbf{K} \mathbf{D}\right)_{w,q}$, independent of the choice of the coordinate system~\cite{WeissCoordinateInvariance,Prost2009}.  

\begin{figure}
  \centering
  \includegraphics[width=0.95\columnwidth]{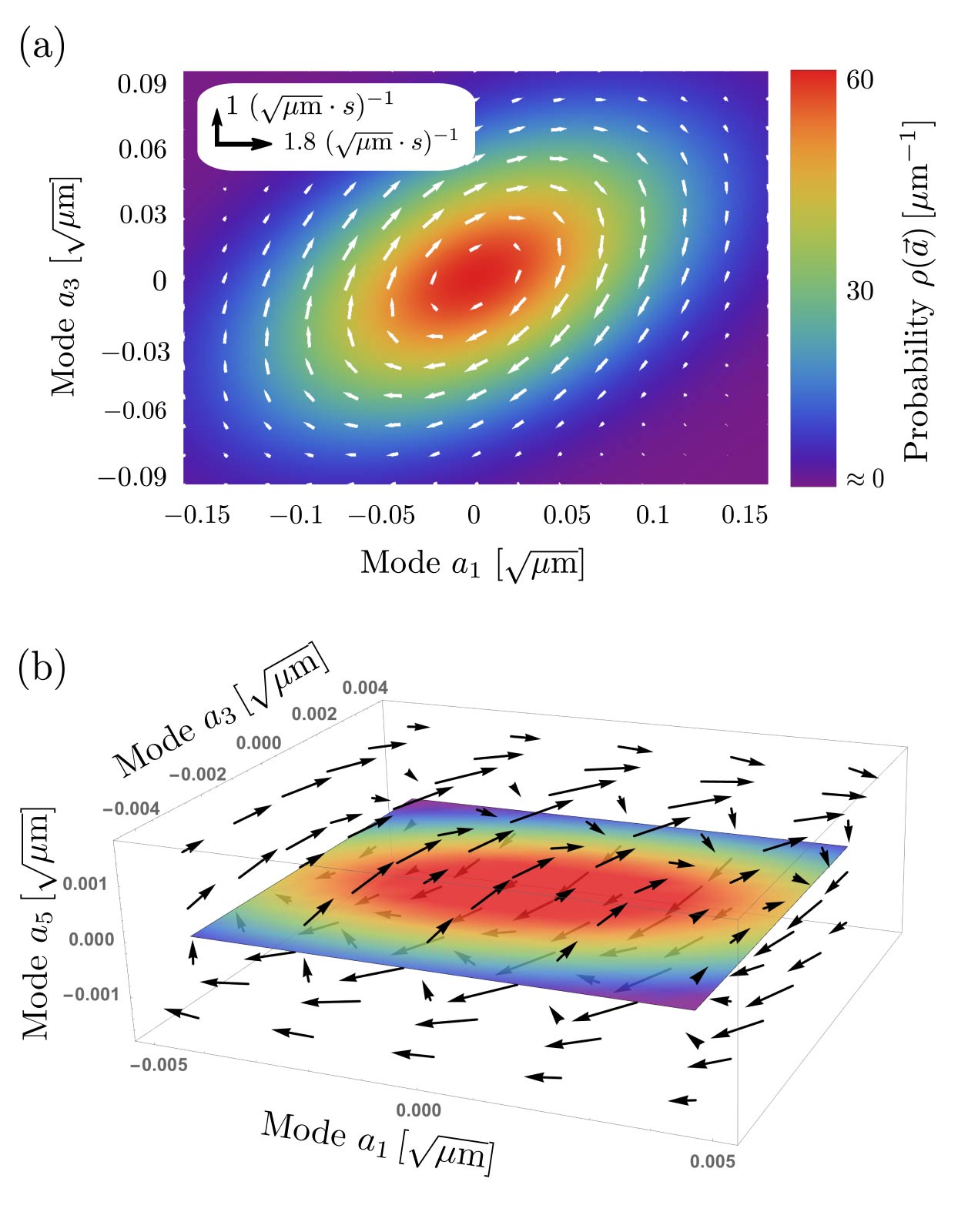}
\caption{(color online) (a) Nonequilibrium probability and current distributions for modes $1$ and $3$. Other degrees of freedom were integrated out. Colors indicate the value of the nonequilibrium probability density $\rho(\vec{a})$ and white arrows indicate the magnitude of probability flux $\vec{j}(\vec{a})$. Parameters were set to model a microtubule in an actin network with parameters as in Fig.~\ref{fig:fluctuationOverModeNumber}, $\tau_\text{M}=0.3$ s, $\ell_\text{M}=1.6\,\mu$m and $f=10$~pN. (b) Nonequilibrium probability ($a_1\times a_3$ plane) and current distributions for modes $1$, $3$ and $5$ where the full current profile was retained. Parameters were chosen as in (a).}
  \label{fig:eqAndNonEqModeDynamics}
\end{figure}

To investigate the behavior of our model, we plotted the steady-state probability and current distributions in the $a_1\times a_3$ plane (Fig.~\ref{fig:eqAndNonEqModeDynamics} (a)), integrating out all other degrees of freedom. The currents (white arrows) exhibit a clockwise circulation in this plane of mode space. Importantly, the breaking of detailed balance in this mode space does not arise from energy exchange between bending modes, but rather from how stochastic motor forces induce mode correlations. In the absence of motor forces, these modes evolve independently, such that the system can be  described by a series of uncoupled 1D systems. Perhaps counter-intuitively, we expect the coupling and resulting currents to vanish in the limit of high motor density (Fig.~\ref{fig:correlationCoefficients}). Interestingly, it has also been argued that heavy-tailed histograms of particle displacements in active networks approach Gaussian distributions in the high motor density limit~\cite{Toyota2011,Gov2015}. Thus, high densities of motors can lead to behaviour resembling thermal equilibrium. In general however, neither Gaussian displacement distributions nor detailed balance in the mode dynamics are sufficient conditions for equilibrium~\cite{Battle2016, Rupprecht}.

It is illuminating to study different projections of the probability currents to characterise the nonequilibrium dynamics. We calculated the joint probability and current distribution in the $a_1\times a_3\times a_5$ subspace, as shown in Fig.~\ref{fig:eqAndNonEqModeDynamics}(b). A slow circulation around the $a_3$ axis is accompanied by a faster circulation around the $a_1$ axis, reflecting the decrease of the relaxation time $\tau_j$ with increasing mode number $j$. The precise structure of the currents is determined by the geometrical details of the interaction of motors with the probe filament, as described by the coupling coefficients $ \mathbf{F}_{q,w}$ (see Fig.~\ref{fig:correlationCoefficients}). 

 A circulating current implies a preferred sense of rotation in configurational space of the underlying stochastic dynamics, with  an associated cycling frequency $\omega_{q,w}$ that scales with the magnitude of the current. To obtain an estimate for this frequency $\omega_{q,w}$, we focus on the origin of the probability current, the active force term $f_{\text{M}}$. From the active mode correlators (Eq.~\eqref{eq:motorModeCorrelation}) we obtain the mode-correlation matrix $\mathbf{C}$ for $ \tau_\text{M}/\tau_q \ll 1$, which we use to solve the Fokker-Planck equation with white-noise. This yields a closed-form of the probability flux $\vec{j}(\vec{a})={\mathbf \Omega} \vec{a}\rho(\vec{a})$, where ${\mathbf \Omega}=(\mathbf{K}+\mathbf{D}\mathbf{C}^{-1})$. Under steady-state conditions, $\vec{j}(\vec{a})$ is purely rotational because $\vec{\nabla}\cdot \vec{ j}=0$, implying purely imaginary eigenvalues of ${\mathbf \Omega}$, which constitute the cycling frequencies associated with the circular probability currents~\cite{WeissCoordinateInvariance}. It is instructive to consider the cycling frequency $\omega_{q,w}$ of a two-dimensional system $a_q \times a_w$  (see SI). The leading-order term of a power expansion of $\omega_{q,w}$ in $\mathbf{F}_{q,w}$ reveals dependencies of the current on relaxation times and coupling coefficients  
\begin{align}
  \label{eq:cyclingFrequencyApprox}
\omega_{q,w}&\approx \frac{ (\tau_q-\tau_w) \mathbf{F}_{q,w}}{\sqrt{\mathbf{F}_{q,q}\mathbf{F}_{w,w}\tau_q\tau_w(\tau_q+\tau_w)^2}}.
\end{align}
The sign of this frequency indicates the direction of the current circulation in mode space (Fig.~\ref{fig:eqAndNonEqModeDynamics}). This underlines that detailed balance in a space of normal mode amplitudes of the filament is broken if and only if motor activity drives several modes with different relaxation times at once such that $\mathbf{F}_{q,w}\neq 0$. Motor activity impacts the filament on the scale $\ell_\text{M}$, inducing a coupling between the bending modes of the filament. This result does not hinge on the motor timescale $\tau_{\rm M}$ and applies even when the driving forces acting at  the discrete points are described by white noise. In contrast, thermal forces impact the filament homogeneously, and thus do not introduce coupling between different bending modes. Interestingly, if we consider an ensemble of filaments dispersed throughout a random network, each filament will sample a different local interaction profile $f_\text{M}(s,t)$.  Therefore, we expect that nonequilibrium mode coupling will vanish when averaging over an ensemble of filaments, restoring detailed balance, even in an active system. 

The cross-correlations and currents predicted by our model could, in principle, be tested experimentally. It is not a priori obvious which projections of mode space will reveal the largest currents, but highly correlated mode pairs constitute likely candidates. Probing detailed balance in fluctuations of probe filaments and measuring cycling frequencies will be an ideal, non-invasive tool to detect and quantify motor activity in biological networks, living cells, and tissues~\cite{BrangwynneNonequilibriumMicrotubule,Guo2014Cell,FakhriHighResolutionMapping}. 
\begin{acknowledgments}
We thank G. Crooks and M. Lenz for discussions. This research was supported by the National Science Foundation under Grant No. NSF PHY11-25915, by the German Excellence Initiative via the program NanoSystems Initiative Munich (NIM) (C.P.B.) and the Deutsche Forschungsgemeinschaft (DFG) Collaborative Research Center SFB 937 (Project A2), the European Research Council Advanced Grant PF7 ERC-2013-AdG, Project 340528 (C.F.S), and then Cluster of Excellence and DFG Research Center Nanoscale Microscopy and Molecular Physiology of the Brain (CNMPB) (C.F.S.). \end{acknowledgments}
% We also thank the Kavli Institute for Theoretical Physics (KITP) at UCSB for hospitality and providing the opportunity for pivotal discussions. 
%References: abbreviations, capitalization

\onecolumngrid
\renewcommand\theequation{S\arabic{equation}}
\setcounter{equation}{0}
\section*{Derivation of the coupling coefficients $\mathbf{F}_{q,w}$ }
\renewcommand\theequation{S\arabic{equation}}
\setcounter{equation}{0}
\noindent In this supplement, we derive the expression for the coupling coefficients  $\mathbf{F}_{q,w}$, which we introduce in the main text. We assume that motor-induced forces interact with the probe filament along its backbone, parametrized by the arc length $s$ at various locations $s_n$. The total force (in units of a force-line density $[\mathrm{N}/\mathrm{m}]$) can therefore be written as
\begin{align}
  \label{eq:totalForce}
    f_\text{M}(s,t) &= \sum\limits_{i}\, f_i \mathcal{T}_i(t)g\left(s-s_i \right)
\end{align}
with $g(s)$ denoting the spatial interaction kernel, which describes the spatial profile of the force-probe interaction.  Motivated by prior work on the deflection of microtubules in actin-myosin networks~\cite{BrangwynneNonequilibriumMicrotubule}, we assume a point-like interaction profile $g(s)\approx \delta(s)$. A projection of $f_\text{M}(s,t)$ onto mode $q$ leads to
\begin{align}
  f_{\text{M},q}(t) &=  \frac{1}{L}\int\limits_{0}^L \mathrm{d}s\, \sum_i \, f_i \mathcal{T}_i(t) y_q(s) \delta(s-s_i) \nonumber \\
  &= \frac{1}{L}\sum_i \, f_i \mathcal{T}_i(t) y_q(s_i). \label{eq:modeMotorForce}
\end{align}

\begin{figure}[h]
  \centering
  \includegraphics[width=0.9\textwidth]{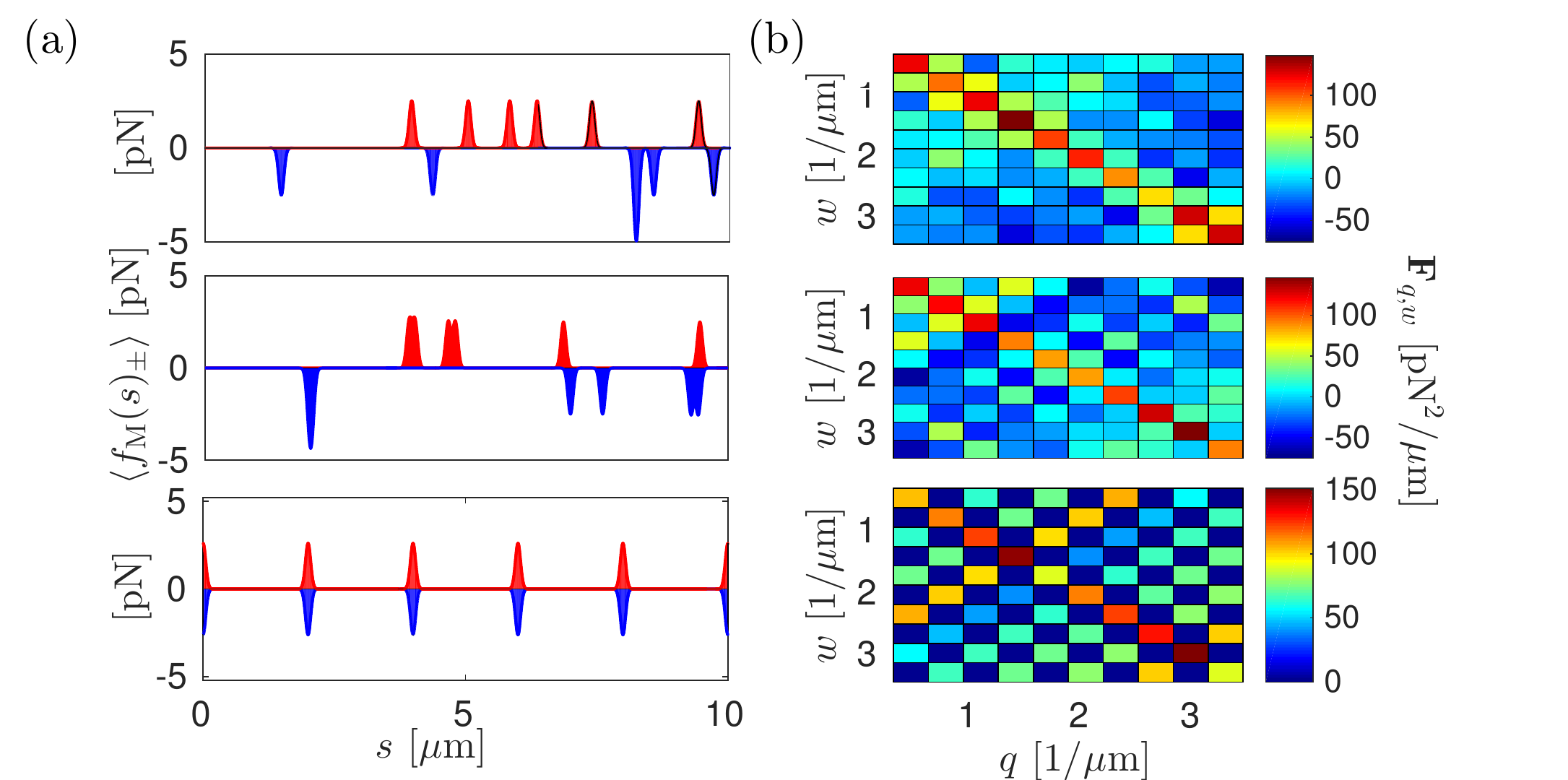}
  \caption{(color online) (a) Temporal average of the sum of positive (red) and negative (blue) components of the motor-induced force $f_M(s,t)$.  In the two top rows, 12 interactions points $s_n$ were chosen from an even distribution $\rho(s_i)=1/L $. Half of these points were assigned negative force coefficients $f_{-}=-f_{+}$. The magnitude of the force was fixed at $f_{+}=|f_{-}|=10$ pN, $L$ was set to $10 \,\mu\text{m}$ and $\tau_\text{off}/\tau_\text{on}=1/3$.  In the bottom row, the interaction points were distributed evenly, such that the average total force and torque vanish. (b) The  mode coupling coefficients $\mathbf{F}_{q,w} $ (first 11 modes) for the distributions shown in panel (a).}
  \label{fig:examplesCouplingCoefficients}
\end{figure}

Assuming that all active processes have  the same internal time scale, and that the processes oat $i$ and $j$ are independent, we can simplify the correlator of active forces $\langle f_q(t)f_w(t')\rangle$:
\begin{align}
  \langle f_{\text{M},q}(t)f_{\text{M},w}(t')\rangle &= \frac{1}{L^2} \sum_i\sum_j f_i f_j y_q(s_i)y_w(s_j) \langle \mathcal{T}_i(t)\mathcal{T}_j(t') \rangle \nonumber \\
  &= \frac{1}{ L^2}\sum_i\sum_j f_i f_j y_q(s_i)y_w(s_j) \delta_{i,j} C_2 e^{-\frac{\lvert t-t'\rvert }{\tau_\text{M}}} \nonumber \\ 
  &= \frac{C_2 e^{-\frac{\lvert t-t'\rvert }{\tau_\text{M}}}}{ L^2}\sum_i f_i^2 y_q(s_i)y_w(s_i) \,  \nonumber
\end{align}
where we used the correlator of the telegraph process, described in Eq.~(2) in the main text. Finally, we can rewrite the sum in the following way, introducing the coupling coefficients $\mathbf{F}_{q,w}=\sum_i f_i^2 y_q(s_i)y_w(s_i)$:
\begin{align}
  \langle f_{\text{M},q}(t)f_{\text{M},w}(t')\rangle &= \frac{1}{L^2} \mathbf{F}_{q,w} C_2 e^{-\frac{\lvert t-t'\rvert }{\tau_\text{M}}}. \label{eq:motorCorrelator}
\end{align}
To illustrate how the distribution of motor interaction points affects these  coefficients,  we plot the matrix of coupling coefficients $\mathbf{F}_{q,w}$ for three different probe-motor interaction geometries $\{s_i\}$ in  Fig.~\ref{fig:examplesCouplingCoefficients}. In the upper two plots the interaction points $s_i$ were drawn from a uniform distribution along the filament $\rho(s_i)=1/L$, while the bottom plot shows the equidistant geometry used in the main text.  

\section*{Derivation of the net cycling frequency $\omega_{q,w}$ in the white-noise regime }
 \noindent  Consider a stochastic trajectory of the vector of mode amplitudes $\vec{a}$, as defined in the main text (Eq.~(8)). In equilibrium, the system transitions stochastically between microstates $\vec{a}$ , obeying detailed balance. In other words, transitions between any two states $\vec{a}_1 \rightleftharpoons \vec{a}_2$ occur at equal rates in both directions. Out of equilibrium, however, there may be transitions that occur more frequently in one , $\vec{a}_1\to \vec{a}_2$, than in the reverse direction. This imbalance of transitions is quantified by the probability current $\vec{j}$. In a nonequilibrium steady-state (NESS), a finite probability current $\vec{j}\neq 0$ must be solenoidal, as any other form of the current would violate the steady-state Fokker-Planck equation $ \vec{\nabla}\cdot \vec{j}=0$ (Eq.~(8) in the main text).
 
Such a solenoidal probability current thus defines a sense of rotation. As an example, consider a two-dimensional configurational phase space in polar coordinates $(r, \theta)$ in a NESS. Here, angular transitions $\Delta \theta$ will occur more frequently in, say, the positive (anti-clockwise) than in the negative (clockwise) direction~$\text{Prob}(\Delta \theta > 0) > \text{Prob}(\Delta \theta < 0 )$. Over time, trajectories will therefore, on average, cycle around the origin in an anti-clockwise sense at a net cycling frequency $\omega=\langle \dot{\theta} \rangle$.

 In the following, we derive an expression for $\omega_{q,w}$ in a reduced two-dimensional space of two modes $a_q$ and $a_w$, where we disregard dynamics of other mode amplitudes. We begin by calculating the closed-form solution of the probability current $\vec{j}(\vec{a})$ from the Fokker-Planck equation 
\begin{align}
  \frac{\partial \rho}{\partial t}(\vec{a},t)&= -\vec{\nabla}\cdot\vec{j}(\vec{a},t) \qquad \text{with}\label{eq:fpe} \\
 \vec{j}(\vec{a},t)&=\mathbf{K}\vec{a}\rho\left(\vec{a},t\right)-\mathbf{D}\vec{\nabla}\rho\left(\vec{a},t\right). \label{eq:current}
\end{align}
Since we are interested in the steady-state, we assume $\partial \rho(\vec{a},t)/\partial t=0$ and $\vec{j}(\vec{a},t)=\vec{j}(\vec{a})$ in Eqs.~(\ref{eq:fpe}) and (\ref{eq:current}) respectively. For white noise, the steady state solution of $\vec{j}(\vec{a})$ can be written as~\cite{WeissCoordinateInvariance} 
\begin{align}
  \vec{j}(\vec{a}) &= \left( \mathbf{K}+\mathbf{D}\mathbf{C}^{-1}\right)\vec{a}\rho\left(\vec{a}\right)\equiv \mathbf{\Omega}\vec{a}\rho\left(\vec{a}\right) \label{eq:closedFormCurrent}
\end{align}
with a multivariate-gaussian probability distribution $\rho(\vec{a})=1/\sqrt{\det(2\pi\mathbf{C})} e^{-\frac{1}{2}\vec{a}^T\mathbf{C}^{-1}\vec{a}}$, and $\mathbf{\Omega}= \mathbf{K}+\mathbf{D}\mathbf{C}^{-1}$.
The correlator $\langle a_q(t)a_w(t')\rangle$ for thermal and coloured active noise is derived in the main text and is given by the sum of the expressions Eq.~(4) and (5), restated below,
\begin{align}
  \langle a_q(t)a_w(t')\rangle_{\text{Th}} &=  \frac{k_B T \tau_q}{L^2\gamma}\delta_{q,w}e^{-\frac{\left | t-t'\right | }{\tau_q}}  \label{eq:thermalModeCorrelation}\\
  \langle a_q(t)a_w(t')\rangle_{\text{M}} &=    \frac{C_2}{L^2 \gamma^2}\mathbf{F}_{q,w}\mathcal{C}_{q,w}\left( t-t'\right). \label{eq:motorModeCorrelation}
\end{align}
We here concentrate on the origin of the probability current, which is the active force term $f_\text{M}(s,t)$ and therefore neglect the thermal equilibrium contribution in Eq.~(\ref{eq:thermalModeCorrelation}) (a more comprehensive analysis can be found in \cite{GladrowPRE}). Furthermore, we take the white-noise limit of Eq.~(\ref{eq:motorModeCorrelation}) for $t=t'$ to obtain the corresponding correlation matrix $\mathbf{C}$ in this limit. To this end, we consider the white-noise limit of the telegraph process $\langle \mathcal{T}_n(t) \mathcal{T}_m(t')\rangle \to C_1 +\frac{C_2}{2}\delta_{n,m}\tau_\text{M}\delta(t-t')$, where the proportionality constant $\tau_\text{M}$ ensures proper scaling of the active force variance. The corresponding limit of the coloured-noise mode correlator Eq.~(\ref{eq:motorModeCorrelation}) can be obtained by expanding $\mathcal{C}_{q,w}( t,t)$ up to linear order in $\tau_\text{M}$, which results in
\begin{align}
  \label{eq:seriesExpActiveCorrelator}
  \langle a_q(t)a_w(t)\rangle_{\text{M}} &\approx \frac{C_2}{L^2 \gamma^2}\mathbf{F}_{q,w}\frac{\tau_q \tau_w}{\tau_q+\tau_w}\tau_\text{M} = \mathbf{C}_{q,w}.
\end{align}
Using the correlation matrix $\mathbf{C}$, we can compute $\mathbf{\Omega}$, and its eigenvalues  
\begin{align}
  \lambda_{\pm}&=   \pm i\frac{  \left (\tau_q -\tau_w\right)\mathbf{F}_{q,w} }{\sqrt{\tau_q\tau_w\left( \left(\tau_q+\tau_w\right)^2 \left( \mathbf{F}_{w,w}\mathbf{F}_{q,q}  \right) -4\tau_q\tau_w \mathbf{F}_{q,w}^2\right) }}.\label{eq:eigenvalues}
\end{align}
The first eigenvalue $\lambda_+$ gives the net cycling frequency, which we wish to derive: $\omega_{q,w}= {\rm Im} (\lambda_{+})$. We note, that the second eigenvalue ${\rm Im} (\lambda_-)=-\omega_{q,w}$ would be the correct frequency, if one exchanged the $a_q$- and $a_w$-axis. 

From the definition of $\mathbf{F}_{q,w}=\sum_n f_n^2 y_w(s_n)y_q(s_n)$ it follows that $\mathbf{F}_{q\neq w}< \mathbf{F}_{q,q}$. We expand the frequency into a power series of the small parameter, the coupling coefficient $\mathbf{F}_{q,w}$ around $0$ and obtain
\begin{align}
  \omega_{q,w}&\approx \frac{\left( \tau_q-\tau_w\right) \mathbf{F}_{q,w}}{\sqrt{\mathbf{F}_{q,q}\mathbf{F}_{w,w}\tau_q\tau_w(\tau_q+\tau_w)^2}}.  \label{eq:approxFrequency}
\end{align}
In principle, this prediction can be tested in experiments. The net cycling frequency can be obtained from sampled trajectories $\{a_q(t), a_w(t)\}$ using $ \omega_{q,w}=\langle \dot{\varphi}_{q,w} \rangle$, where $\varphi_{q,w}(t)= \tan^{-1}\left( a_w (t)/ a_q(t) \right)$ is the polar angle of $a_q(t)$ and $a_w(t)$, such that  $\omega_{q,w}= \langle \left(\dot{a}_wa_q-\dot{a}_qa_w\right)/\left( a_q^2+a_w^2\right)\rangle$. 

 % by adding up angular changes of the stochastic trajectory in the $q$-$w$ plane, $\Delta \theta_{i}$ over time and dividing by the total time of measurement $t_{\rm total}$, $\omega_{q,w}= \sum_i \Delta \theta_i/t_{\rm total}$.

\end{document}